\documentclass[fleqn,usenatbib]{mnras}

\usepackage{newtxtext,newtxmath}

\usepackage[T1]{fontenc}
\usepackage{ae,aecompl}

\usepackage{graphicx}	
\usepackage{amsmath}	
\usepackage{amssymb}	
\usepackage[flushleft]{threeparttable}

\title[Influence of magnetic field in the CNM]{The influence of magnetic field on the CNM mass fraction and its alignment with density structures}

\author[M. A. Villagran and A. Gazol]{
M. A. Villagran,$^{1}$\thanks{E-mail: m.villagran@irya.unam.mx}
A. Gazol,$^{1}$
\\
$^{1}$Instituto de Radiastronom\'ia y Astrof\'isica, Universidad Nacional Aut\'onoma de M\'exico, 58089 Morelia, Michoac\'an, M\'exico\\
}

\date{Accepted XXX. Received YYY; in original form ZZZ}

\pubyear{2017}

\begin{document}
\label{firstpage}
\pagerange{\pageref{firstpage}--\pageref{lastpage}}
\maketitle

\begin{abstract}
To contribute to the understanding of the magnetic field's influence on the segregation of CNM in the solar neighbourhood we analyse MHD simulations which include the main physical characteristics of the local neutral atomic ISM. The simulations have a continuous solenoidal Fourier forcing in a periodic box of 100\,pc per side and an  initial uniform magnetic field ($\vec{B_0}$) with intensities ranging between $\sim 0.4$\,$\mu$G and $\sim 8$\,$\mu$G. Our main results are: i) the CNM mass fraction diminishes with the increase in magnetic field intensity. ii) There is a preferred alignment between CNM structures and $\vec{B}$ in all our $B_0$ range but the preference weakens as $B_{0}$ increases. It is worth noticing that this preference is also present in two-dimensional projections  making an extreme angle ($0$ or $\pi / 2$) with respect to $\vec{B_0}$ and it is only lost for the strongest magnetic field when the angle of projection is perpendicular to $\vec{B_0}$.
iii) The aforementioned results are prevalent despite the inclusion of self-gravity in our continuously forced simulations with a mean density similar to the average value of the solar neighbourhood. iv) Given a fixed $B_0$ and slightly higher mean densities, up to double, the effects of self-gravity are still not qualitatively significant. 
\end{abstract}

\begin{keywords}
(magnetohydrodynamics) MHD -- ISM: magnetic fields -- ISM: structure -- (Galaxy:) local interstellar matter
\end{keywords}




\section{Introduction}\label{introduction}
The neutral atomic component of the interstellar medium (ISM) is characterized by a multiphase behaviour, \citep[see][]{FieldHG69, Wolfire1995}.
In fact, due to the isobaric mode of the thermal instability (TI) there is a possibility for the gas to segregate in two thermally stable components, the cold neutral medium (CNM) and the warm neutral medium \citep[WNM, ][]{Field65}. The TI is originated by the balance between cooling and heating processes due to the interaction of the background diffuse radiation and cosmic rays with the gas and dust present in the atomic ISM \citep[see for example][]{Wolfire2003}.  This segregation process is coupled with the rest of the physical phenomena that affect the ISM, for example: turbulence introduced at small scales by supernovas and at larger scales by the dynamics of the Galactic disk, the gravitational potential coming from the Galaxy, the self-gravity of the gas, and magnetic fields at different scales.

The atomic neutral gas in the Galaxy has been extendedly observed, most commonly using absorption and emission of the 21 cm hydrogen line, as recalled on the reviews by \cite{Kulkarni1987}, \cite{Dickey1990}, \cite{Dickey2002}, and \cite{Kalberla2009}. The Millennium Arecibo 21 Centimeter Absorption-Line Survey, described in \cite{Heiles2003}, has allowed the characterization of the physical properties of the CNM gas in the solar neighbourhood, which is the region of interest for the present work. However, despite of the large observational and numerical effort there are still unresolved questions about the CNM. Some examples of this are: what fraction of neutral gas is in CNM? how is the CNM distributed? how is the mixture of CNM and WNM in different physical environments? and, what is the relationship between atomic gas phase transition and star formation? In fact, as mentioned in \cite{GASKAP}, one of the goals of the Square Kilometre Array (SKA) is to complete a Galactic survey (GASKAP) of HI and OH that will give a deeper insight to the properties of the HI in the Milky Way.

One of the quantities used to study the mixture between CNM and WNM is the CNM mass fraction. This fraction is usually measured as the ratio of CNM column density to total HI column density along individual lines of sight, \linebreak \citep[e.g. ][]{Heiles2003, Roy2013, Murray2015, Roy2017}. In the solar neighbourhood, the cold gas mass fraction distribution shows a  peak around $\sim 0.4$, \citep[][]{Heiles2003}.  This value has been confirmed by applying other techniques by \cite{Pineda2013}, who used the 158\,$\mu$m CII line observations obtained with the Herschel/HIFI instrument and combined them with HI (21 cm), $^{12}$CO and $^{13}$CO (J = 1 $\rightarrow$ 0 transitions) maps. From the theoretical point of view, explaining the behaviour of this fraction along the Galactic disk is still a challenge, \citep[see e.g. ][]{US}, however this represents an important constraint to ISM models \citep[][]{Wolfire2015}. Observationally, this quantity is sometimes needed to analyse data. For example, \cite{Herrera2017} use the CNM mass fraction to estimate the thermal pressure in the cold gas in galaxies from the Hershel/KINGFISH sample through CII 158\,$\mu$m line observations combined with HI and CO maps. They find that varying the fraction from 0.3 to 0.7 gives discrepancies in pressure up to a factor of 2.3. This puts in evidence the need to narrow the CNM mass fraction values given the local physical characteristics.

Moreover, observations of the neutral atomic ISM have revealed that magnetic fields should play an important role in the gas dynamics as the magnetic energy density seems to be comparable to the turbulent one and it dominates over the thermal energy density. The aforementioned results together with a brief description of the techniques used to observe the magnetic field on the dense ISM and with its general properties can be found in \cite{Heiles2005}. Analytically, the influence of the magnetic field in the linear regime of the TI has been studied since \cite{Field65}, where three possible effects are mentioned: the presence of the two polarizations of Alfv\'en waves propagating along $\vec{B}$, the possible inhibition of the TI when the perturbations propagate perpendicular to the magnetic field, and the reduction of the thermal conductivity across $\vec{B}$. From the numerical perspective, the interplay between magnetic fields and the TI has been studied for a couple of decades, with a variety of approaches and objectives in mind. Some of the main topics addressed have been: the interaction of neutral ISM with non-linear MHD waves \citep[e.g.][]{Elmegreen1997, Hennebelle2006}, the influence of $\vec{B}$ in the resulting structure in a supernovae driven ISM \citep[e.g.][]{Mac2005, Avillez2005, GentII_2013, Hill2012, Kim2015, Evirgen2017, Pardi2017}, the effects of $\vec{B}$ on the structure resulting from collision of WNM flows \citep[e.g.][]{Hennebelle2000, Inoue2008, HennebelleAndEnrique, Heitsch2009, Inoue2009, Banerjee2015, Fogerty2016}, and the relative orientation between diffuse density structures and magnetic field \citep[][]{Inoue2016}. The last work mentioned was motivated by recent observations concerning the possible alignment between diffuse structures and magnetic field.  One of those observational results is presented in \cite{Clark2014}, where the authors apply the Rolling Hugh Transform algorithm to observations of diffuse HI regions obtained from the GALFA-HI and the Parkes Galactic All Sky surveys. They isolate structures with linear properties and find an alignment between the magnetic field and these fibres but, nevertheless, with an ample dispersion. Additionally, the maps obtained by the Planck mission have been used to trace galactic matter structures (dominated by CNM) and the polarization of light due to dust grains associated to these structures. In \cite{PlanckXXXII} the orientation between the polarizing magnetic field and the CNM structures is studied using linearly polarized emission from dust at 353 GHz. They find a preferred alignment between $\vec{B}$ and CNM dominated dense structures and an apparent breach of this correlation for column densities corresponding to molecular structures.

The advent of this kind of observational work, together with the uncertainties concerning the spatial distribution of the CNM in the Milky Way, wakes the necessity for numerical simulations to supplement the current knowledge of the diffuse ISM and the relationship it has with the Galactic magnetic field. The purpose of the present paper is thus to evaluate and quantify the effects that magnetic fields have on the segregation of the neutral atomic ISM due to TI on a constantly agitated medium. Having in mind the general questions formulated in the previous paragraphs together with the reported values for the physical conditions of the solar neighbourhood, we calculated the CNM mass fraction resulting from MHD simulations that differ between them on the intensity of the initial magnetic field that permeates the gas. We have also searched for the angular correlation between the magnetic field and the cold dense structures formed in our simulations. Additionally, we studied the relevance of self-gravity analysing models with densities higher than the mean value in the solar neighbourhood. 

The distribution of the paper is as follows. First, in section \ref{methodology} we describe the methodology used and all the initial conditions of our simulations. Then, in section \ref{results} we start by portraying the basic physical characteristics of simulations intended to mimic the physical conditions of the solar neighbourhood. This section also presents the results we obtained, both for the cold gas mass fraction and the alignment of dense gas with $\vec{B}$. Further, section \ref{discussion} contains a discussion on our results. Finally, we give our conclusions in section \ref{conclusion}.


\section{Methodology}\label{methodology}

In this article we analyse eighteen simulations divided in two sets; the "main set", used to study the effects of varying the magnetic field intensity in a medium with other physical conditions similar to those of the solar neighbourhood, and the "high density set", which is used to test the effects of self-gravity in a denser medium. Our models result from solving the three-dimensional magneto-hydrodynamic equations with periodic boundary conditions using an Eulerian scheme in a Cartesian grid, and are similar to the models presented in \cite{US}. Using $512^3$ cells for a $100$\,pc per side, the simulations have a resolution of $\sim 0.2$\,pc. Artificial energy injection is achieved by solenoidal Fourier forcing at a fixed wave-number, corresponding to $50$\,pc, and with a constant energy injection rate. The supply of non localized energy can introduce artificial effects \citep[see ][]{Gazol2005}, but nevertheless allows us to control the $v_{rms}$ speed as well as the characteristic scale of energy injection, which results on models reproducing the main physical properties of the atomic ISM in the solar neighbourhood (see \ref{properties}). Note that at the scale that we use, the effects of the galactic HI disk rotation and stratification can be ignored, \citep[see ][]{Kalberla2009}.

We start the simulations with a uniform gas at rest and in thermal equilibrium. In our main set of simulations the initial density is $n_0 = 2$\,cm$^{-3}$ for which the corresponding thermal equilibrium temperature, according to our cooling function, is 1500\,K. These density and temperature values are similar to the ones of the solar neighborhood \citep[see ][]{Wolfire1995}, and lie on the unstable range according to our thermal equilibrium. Our set of denser simulations has either $n_0 = 3$\,cm$^{-3}$ or $n_0 = 4$\,cm$^{-3}$, which are also unstable densities with respect to our thermal equilibrium. Our cooling function is a piece-wise power law fit with a constant heating of $\Gamma = 22.4 \times 10^{-27}$\,erg\,cm$^{3}$\,s$^{-1}$, where both the cooling and heating functions are determined based on the values presented in \cite{Wolfire2003}. The cooling and heating functions used in this paper have been presented before in \cite{US}. Our magnetohydrodynamic simulations have an initial uniform magnetic field along the $x$ direction with the intensities $B_0$ chosen to be around the values reported for the solar neighborhood ($\sim 6 \mu G$) and for our region on the galactic disk ($\sim 3 \mu G$), given by \cite{Beck2008} and \cite{Jansson2012}, respectively. The initial fields are presented in Table~\ref{tab:B0_values} and Table~\ref{tab:B0_dense} for the main set and the denser simulations, respectively.

Each set of initial conditions has been used in a simulation without self-gravity (S) and in one with it (G). The main set consists of five pairs of simulations, four of which have magnetic field and one which is purely hydrodynamical. The high density consists of four pairs of simulations, two MHD and two hydrodynamical with different $n_0$. The way of forcing and the energy injection rate is identical between simulations in the main set. For the high density set we adjusted the energy injection rate to obtain a $v_{rms}$ similar between all the simulations. According to the initial magnetic field and the value of $v_{rms}$ (Table~\ref{tab:B0_values} and Table~\ref{tab:B0_dense}, fifth columns) the simulations can be classified as sub-Alfv\'enic (B01 and B05), trans-Alfv\'enic (B10, B10n3, B10n4) and super-Alfv\'enic (B20). Here we have used the terms sub/trans/super Alfv\'enic freely, as the initial conditions have a $v_{rms} = 0$.

\begin{table}
	\centering
	\caption{Main set, physical run properties.}
	\label{tab:B0_values}
    \begin{threeparttable}
	  \begin{tabular}{lccccr}
		  \hline
		  Sim.\tnote{1} & $B_0$\tnote{2} & $\beta_0$\tnote{3} & v$_A|_{B_0}$\tnote{4} & $v_{rms}$\tnote{5} & Gravity\\
		  \hline
          name & $\mu$G &         & km s$^{-1}$ & km s$^{-1}$ & \\
          \hline
          B00S & 0.0 & - & 0.0 & 7.8 & Off\\
		  B01S & 0.4 & 60 & 0.64 & 7.2 & Off\\
		  B05S & 2.1 & 2.4 & 3.2 & 6.8 & Off\\
		  B10S & 4.2 & 0.60 & 6.4 & 6.7 & Off\\
          B20S & 8.3 & 0.15 & 13 & 6.5 & Off\\
          B00G & 0.0 & - & 0.0 & 7.7 & On\\
		  B01G & 0.4 & 60 & 0.64 & 7.1 & On\\
		  B05G & 2.1 & 2.4 & 3.2 & 6.8 & On\\
		  B10G & 4.2 & 0.60 & 6.4 & 6.8 & On\\
          B20G & 8.3 & 0.15 & 13 & 6.5 & On\\
		  \hline
	 \end{tabular}
     \begin{tablenotes}
       \item[1] All of them with $n_0 =2$\,cm$^{-3}$.
       \item[2] Initial magnetic field.
       \item[3] $\beta$ at initial magnetic and thermal pressure.
       \item[4] Alfv\'en speed calculated with $B_0$.
       \item[5] Time averaged value.
     \end{tablenotes}
   \end{threeparttable}
\end{table}

\begin{table}
	\centering
	\caption{High density set, physical run properties.}
	\label{tab:B0_dense}
    \begin{threeparttable}
	  \begin{tabular}{lccccr}
		  \hline
		  Sim.\tnote{1} & $B_0$\tnote{2} & $\beta_0$\tnote{3} & v$_A|_{B_0}$\tnote{4} & $v_{rms}$\tnote{5} & Gravity\\
		  \hline
          name & $\mu$G &         & km s$^{-1}$ & km s$^{-1}$ & \\
          \hline
          B00Sn3 & 0.0 & - & 0.0 & 7.9 & Off\\
          B00Sn4 & 0.0 & - & 0.0 & 7.9 & Off\\
		  B10Sn3 & 4.2 & 0.44 & 5.2 & 6.8 & Off\\
		  B10Sn4 & 4.2 & 0.41 & 4.5 & 6.9 & Off\\
          B00Gn3 & 0.0 & - & 0.0 & 7.7 & On\\
          B00Gn4 & 0.0 & - & 0.0 & 7.7 & On\\
		  B10Gn3 & 4.2 & 0.44 & 5.2 & 6.8 & On\\
		  B10Gn4 & 4.2 & 0.41 & 4.5 & 6.9 & On\\
		  \hline
	 \end{tabular}
     \begin{tablenotes}
       \item[1] The name of the simulation indicates its initial density.
       \item[2] Initial magnetic field.
       \item[3] $\beta$ at initial magnetic and thermal pressure.
       \item[4] Alfv\'en speed calculated with $B_0$.
       \item[5] Time averaged value.
     \end{tablenotes}
   \end{threeparttable}
\end{table}


\section{Results}\label{results}

Our main interests are the mass fraction of cold segregated gas and the relative orientation between the density structures and the magnetic fields for each simulation, both in three dimensions and in two different two dimensional projections. Prior to studying these quantities we first need to verify that our chosen parameters for the main set result in gas with statistical properties akin to the ones of the ISM in the solar neighbourhood.

\subsection{Main set physical properties}\label{properties}

Some useful parameters comparable with observations are the different kinds of energy , the sonic and Alfv\'enic Mach numbers and the plasma beta parameter.

\subsubsection{Mean energies}

\begin{figure}
	\includegraphics[width=\columnwidth]{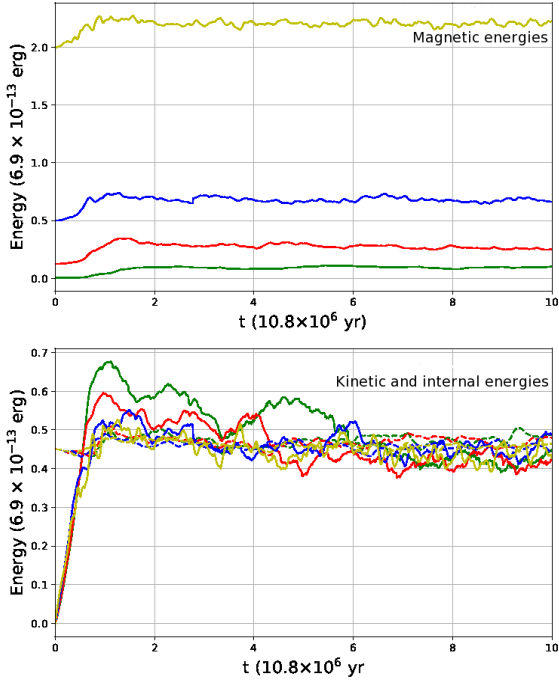}
    \caption{Evolution of the energies, based on the code units, for simulations without self-gravity. Different colours for the magnetic (top), the kinetic (bottom, solid line) and the thermal energy (bottom, dashed line) are for B01 (green), B05 (red), B10 (blue) and B20 (yellow).}
    \label{fig:energiesS}
\end{figure}

The evolution of the kinetic, internal and magnetic energies resulting form our non-self-gravitating magnetized simulations is displayed in Fig.~\ref{fig:energiesS}. The time scale used in these plots is based on the code's time unit $t=1.08 \times 10^6$\,yr. These energies show a smooth evolution and reach a stationary regime. This expected behaviour allows for a statistical analysis to be performed with a significant amount of data. From the same figure it can be noticed that in all cases the kinetic and the internal energy are comparable, whereas the magnetic energy is comparable (B05 and B10), lies above (B20) or bellow (B01) the other energies depending on $B_0$. For each value of $B_0$ the evolution of the energies for the self-gravitating simulations (not shown) is similar.

\subsubsection{Mach numbers}

\begin{figure}
	\includegraphics[width=\columnwidth]{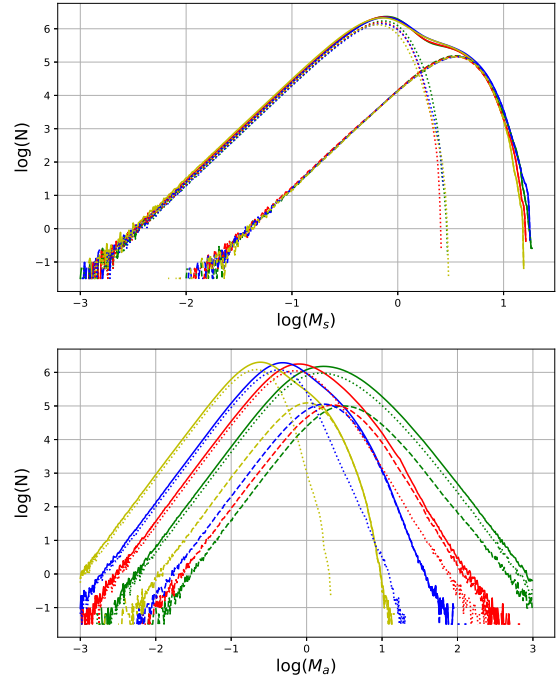}
    \caption{Time averaged histograms of $M_s$ (top) and $M_a$ (bottom). The colour code is the same as in Fig~\ref{fig:energiesS}. In each panel solid, dashed and dotted lines are for all the gas, cold gas, and warm gas, respectively.}
    \label{fig:Mach_Histograms}
\end{figure}

The curves plotted in Fig.~\ref{fig:Mach_Histograms} are the histograms of both the sonic (top) and Alfv\'enic (bottom) Mach numbers for the simulations without self-gravity in the main group. In both panels the solid lines are for all of the gas in the box, the dashed lines represent the cold component ($T \leq 278.57$\,K) histogram, and the dotted lines are for the warm component ($T \geq 5593$\,K).  The temperature $T = 278.57$\,K ($T = 5593$\,K) represents the upper (lower) value of the cold (warm) stable branch in our cooling function.

The sonic Mach number, M$_s$, histograms are double peaked as expected due to the contribution of both stable phases of the atomic ISM; with the sub-sonic peak corresponding to the WNM and the super-sonic one to the CNM.
Our histograms for the cold gas's $M_s$ have their maximums at $\sim 3.47$ (B01 and B05), $\sim 3.72$ (B10) and $\sim 3.63$ (B20). These values are surprisingly close to the typical value for the sonic Mach number of the CNM in the solar neighbourhood, which is $\sim 3.7$ \citep[see ][]{Heiles2003}.

The Alfv\'enic Mach number, $M_a$, histograms are single peaked, but there is nevertheless a noticeable difference between the peaks of the WNM and those of the CNM. For all of the simulations the WNM peak is on the sub-Alfv\'enic region except for B01. At the same time, for all the simulations the CNM peak is in the super-Alfv\'enic region, $\sim 3$ (B01), $\sim 2.3$ (B05) and $\sim 1.7$ (B10),  besides B20, which has its peak on the trans-Alfv\'enic region with $M_a \sim 1$. Here, again, the distributions resulting from the self-gravitating simulations are similar to the ones presented above. Observationally, it is reported in \cite{Heiles2003} a value $M_{a}^2 = 1.3$ ($M_{a} \sim 1.14$) in the local ISM for CNM with a temperature $T \sim 50$\,K associated with a mean magnetic field $B \sim 6$\,$\mu$G, which is between the values we find for B10 ($B_0 \sim 4$\,$\mu$G) and B20 ($B_0 \sim 8$\,$\mu$G).

\subsubsection{Beta parameter}

\begin{figure}
	\includegraphics[width=\columnwidth]{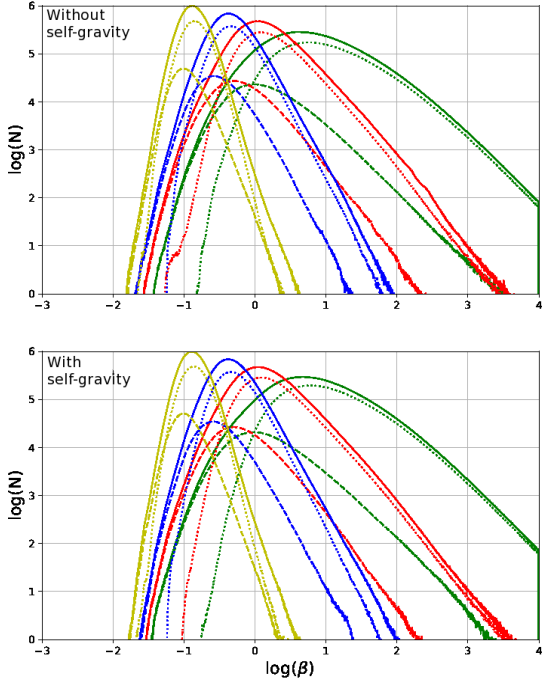}
    \caption{Time averaged $\beta$ histograms for simulations without (top) and with (bottom) self-gravity from the main set. The colour code is the same as in Fig~\ref{fig:energiesS} and the line code is the same as in Fig~\ref{fig:Mach_Histograms}.}
    \label{fig:3D_betas}
\end{figure}

In Fig.~\ref{fig:3D_betas} we show the time averaged histograms of the $\beta$ parameter for all the magnetized simulations.  The main difference between the simulations with self-gravity and the ones without it, is a small change in the slope at the $log(\beta) < 0$ region of the curves, yet apart from that they seem quite similar. When we take into account all of the gas (solid lines), two regimes can be observed in both sets of simulations, depending on where does the maximum of the histograms lie with respect to $log(\beta) = 0$. On one hand, the histograms resulting from sub-Alfv\'enic simulations, have a maximum value in the region $log(\beta) \leq 0$, are narrower, and present less change in slope between the simulations with and without self-gravity. On the other hand, the super-Alfv\'enic and trans-Alfv\'enic simulations show histograms considerably broader and with a slope that is more susceptible to the presence of self-gravity. For the warm component (dashed lines) the trend is the same as for all of the gas. For the cold gas (dotted lines), the maximum of its $\beta$ distribution falls in the region $log(\beta) \leq 0$ in all of the simulations, meaning that in the cold gas the magnetic pressure dominates over the thermal pressure independently of the initial magnetic intensity or the presence of self-gravity.

The histograms becoming wider as $B_0$ decreases is a natural consequence of the fact that the width of the thermal pressure distribution does not have  considerable variations for different $B_0$ values, while for simulations with low initial magnetic field  the magnetic pressure can attain both, extremely low and relatively high values (compared with its initial value) due to the dragging of the magnetic field  by gas motions. For the same reason, the distance between the peaks of all gas histograms and the $\beta$ parameter at initial conditions,  $\beta_0$, decreases with $B_0$ (Table~\ref{tab:B0_values}, column 3). On the other hand, the higher shift towards low $\beta$ values for the cold gas suggests an increase of the magnetic field intensity in a large fraction of cold cells. Although in the low $B_0$ case the cold gas histogram peaks barely below zero, the distance between this peak and the one of the warm gas histogram suggests a larger magnetic field intensity contrast for this case.

The CNM has been classified as a gas with a small plasma beta \citep[][]{Heiles2005}. This agrees with our $\beta$ histograms for the cold gas, which have maximums in low betas ranging between 0.098 (B20S) and 0.93 (B01S). The corresponding $\beta$ for a background magnetic field of $6$\,$\mu$G should lie between our B10 ($\beta \sim 0.26$) and B20 simulations, thus, reinforcing the point of using our selected initial conditions in the main set.

\subsection{CNM gas fraction}\label{sec:fraction}

An important consequence of the development of the TI is the formation of cold gas. Being so, it is relevant to measure the amount of resulting cold gas at the stationary regime established in our simulations. In particular, the mass fraction can be compared with recent observations. With numerical data cubes the most immediate measurable fraction is the three-dimensional cold gas mass fraction (f$_{3d}$). Moreover, we estimated two-dimensional fractions using projections of our data considering both, all the lines of sight (f$_{2d}$) and only lines of sight with cold gas in them (f$_{2d0}$). A proper definition of these fractions is given bellow.

\subsubsection{Main set: effects of varying $B_0$}\label{sec:fractions_main_set}
\begin{figure}
	\includegraphics[width=\columnwidth]{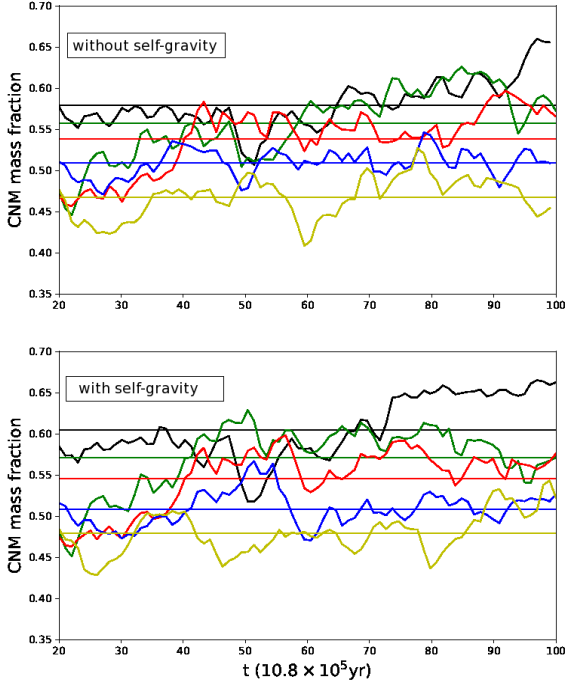}
    \caption{Evolution of the cold gas mass fraction resulting from the three-dimensional data of the main set, without self-gravity (top) and with self-gravity (bottom). The straight lines represent the average values, the black lines represent the simulations B00 while the rest of the colour scheme is the same as the one used in Fig.~\ref{fig:energiesS}.}
    \label{fig:3D_fraction}
\end{figure}

The evolution of the resulting f$_{3d}$s and their average values, for our main set, are displayed in Fig.~\ref{fig:3D_fraction}, where the average values, represented by straight lines, were calculated during the stationary regime in a period that begins at $t\sim 4.4 \times 10^7$\,yr and ends with the simulation's last point at $t\sim 1.1 \times 10^{8}$\,yr, based on the code's time unit, and are displayed in the second column of Table~\ref{tab:all_fractions}.

As expected, B00 simulations show a difference in average CNM fraction, being bigger for the self-gravitating case. On the other hand, when we look only at MHD runs, it is clear that an increase in $B_0$ lowers f$_{3d}$ independently of the presence of self-gravity. However when looking in more detail, it can be seen that the presence of self-gravity has different effects for different values of $B_0$. More specifically, the runs B05 and B10 show marginal differences between the self-gravitating and the non-self-gravitating case, whereas the average CNM mass fraction is lowered for the simulations with the strongest initial magnetic field and raised for the ones with the weakest value of $B_0$ when self-gravity is present. Two things are in play here: self-gravity needing larger enough dense structures to be relevant and large magnetic fields partially inhibiting the TI. Our simulation domain is highly gravitationally stable. For the main set, there is a Jean's length $\lambda_J\sim 552$\,pc at initial conditions. The models with higher initial density (with smaller $\lambda_J$) are discussed at the end of this section. At thermal equilibrium $\lambda_J\sim 27$\,pc for $n \sim 50$\,cm$^{-3}$. Thus, for gravity to have an important role in bounding structures together and making them more massive we need to form relatively large dense regions via thermal instability.\footnote{The maximum density we can resolve, based on the criterion given in \cite{Truelove1997}, is $n \sim 6350$\,cm$^{-3}$. This density is not achieved in our simulations with magnetic fields. It is reached in our hydrodynamic simulations with self-gravity, however the low amount of cells that achieve this makes it a non-significant issue. A possible effect of having collapsed points in a box with a fixed amount of mass is that the WNM gas could be misestimated, but this lies beyond the scope of the present work.}  On the other hand, according to \cite{Field65} there is a critical magnetic field, $B_{crit}$, above which stronger magnetic fields inhibit the development of the TI for perturbations propagating in a direction normal to $\vec{B}$. For our cooling function, the expresion for $B_{crit}$ \citep[eq.(49) in ][]{Field65} simplifies to $B_{crit} = c_s [4 \pi n \gamma^{-1} \beta_i (1 - \beta_i)]^{1/2}$ \citep[here the notation is the same as the one used in][]{US} which implies that for the initial conditions of our simulations is $B_{crit} \sim 1.1$\,$\mu$G, while for the temperatures delimiting the thermally unstable regime  $B_{crit} \sim 1.33$\,$\mu$G and  $\sim 0.45$\,$\mu$G, at the warm and cold limit, respectively. From these values we see that the simulations B01 can segregate gas freely and generate dense regions that may be bounded by the effects of self-gravity, while the TI in the B20 simulations has a higher probability of being partially inhibited by suppressing $B > B_{crit}$ (higher $B_0$ that is hardly perturbed), possibly preventing the formation of large enough high-density regions.

In a previous work, \cite{US}, we studied how varying the $v_{rms}$ speed affects the cold mass gas fraction in purely hydrodynamic simulations with a similar cooling function as the one used here. We analysed simulations with a wide range of values from $v_{rms} \sim 4$\,km\,s$^{-1}$ to $v_{rms} \sim 17$\,km\,s$^{-1}$ yet the maximum difference it made in mean f$_{3d}$ values was $\sim 9 \%$. This is not as significant as the difference obtained here with a varying magnetic field as in this case we have a $\sim 16 \%$ maximum difference. This fact suggests that the variations in the magnetic field around realistic values for the local ISM are more important for the atomic gas segregation than the variations in the $v_{rms}$.

\begin{figure}
	\includegraphics[width=\columnwidth]{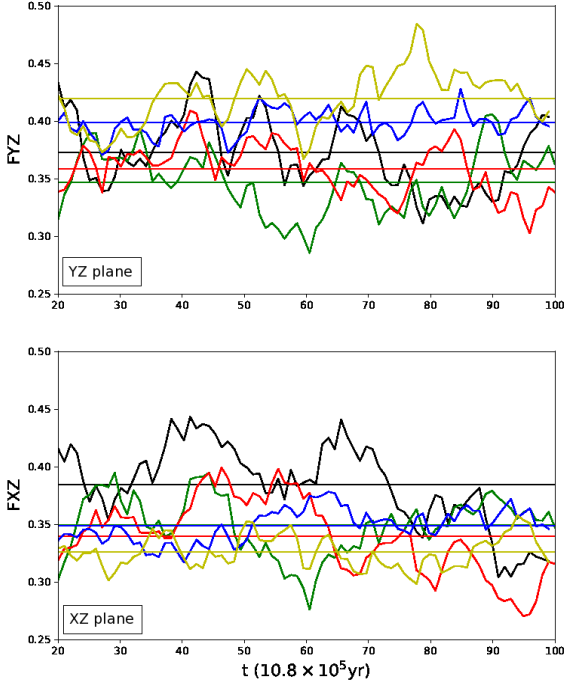}
    \caption{Evolution of f$_{yz}$ (top) and f$_{xz}$ (bottom) for the cold gas resulting from non self-gravitating simulations on the main set. The straight lines represent the average values and the colours are the same as the ones used in Fig.~\ref{fig:3D_fraction}.}
    \label{fig:2D_fractionsS}
\end{figure}

\begin{figure}
	\includegraphics[width=\columnwidth]{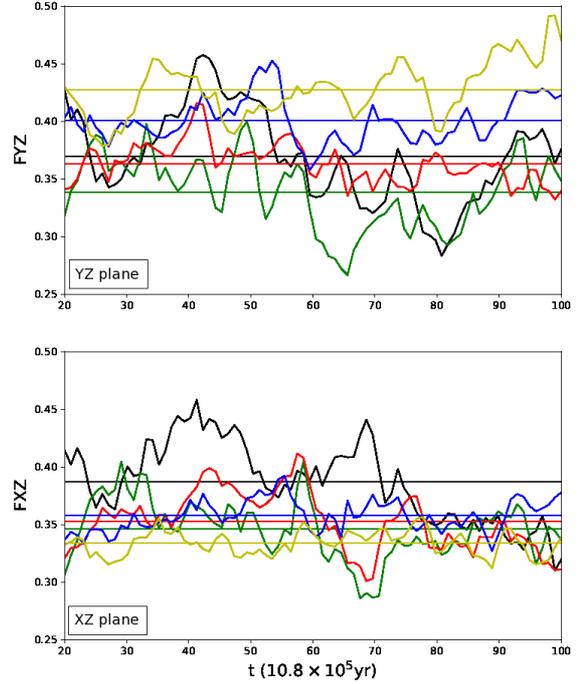}
    \caption{Evolution of f$_{yz}$ (top) and f$_{xz}$ (bottom) for the cold gas resulting from self-gravitating simulations on the main set. The straight lines represent the average values and the colours are the same as the ones used in Fig.~\ref{fig:3D_fraction}.}
    \label{fig:2D_fractionsG}
\end{figure}

The values of f$_{3d}$ presented above result from our three dimensional data, but the projection effects due to the geometry of the cold structures may be lost in that analysis. To address this, similar calculations were done for two-dimensional projections in the YZ and the XZ planes, f$_{2d}$ (f$_{2d0}$). We compute the corresponding CNM mass fractions, f$_{yz}$ (f$_{xz}$) and f$_{yz0}$ (f$_{xz0}$), as described in eqs.(5) and (6)  in \cite{US}. Note that f$_{yz}$ is the equivalent to f$_x$ in \cite{US}. In the aforementioned work, eq.(5) was used to calculate the CNM mass fraction in a single column, i.e. it is the ratio of CNM mass to total gas mass in a single ''line of sight''. This is repeated for the full $512^2$ columns in our planes resulting in a list of CNM fractions on individual columns. Following this, eq.(6) is used to calculate the mean value of the list of CNM fractions. The chosen planes are extreme in the angles they make with the original configuration of $\vec{B_0}$, which is normal to the YZ plane and lies parallel to the XZ plane. In Figs.~\ref{fig:2D_fractionsS} and ~\ref{fig:2D_fractionsG} we show the evolution of f$_{yz}$ and f$_{xz}$ for the simulations, from the main set, without and with self-gravity, respectively. There are several things important to mention about these results. First of all, for the complete set of simulations the average fractions in both projections (Table~\ref{tab:all_fractions}, third and fourth columns) have lower values than their three-dimensional counterparts, which is a natural consequence of the way in which projected fractions are computed. In fact f$_{3d}$ can be interpreted as the column density weighted mean value of the mass fraction along directions perpendicular to the plane, while the corresponding projected mass fraction is just the arithmetic mean of the same values. This means that f$_{yz}$ and f$_{xz}$ do not take into account the fact that mass fractions from directions with higher column densities contribute more to the mass fraction. A thing to notice is that for f$_{2d}$ the influence of varying $B_0$ is less noticeable on the XZ plane, where the maximum difference in the average mass fraction is $\sim 3\%$, while on the YZ plane the difference is as high as $\sim 8 \%$ for the non-self-gravitating magnetized simulations. In the self-gravitating case differences of $\sim 3 \%$ (XZ) and $\sim 10 \%$ (YZ) are seen. Even if the average value of f$_{2d}$ for the non-magnetized simulations is similar between projections, both in the simple case and in the self-gravitating one, there is a noteworthy qualitative difference on the effects of the presence of magnetic field. While for f$_{xz}$, similar to the fraction resulting from the 3D case, the non magnetized average is above all others, for f$_{yz}$ it is between the strong and the weak magnetic intensities, indicating important geometric differences between the cold gas structures resulting from initially sub Alfv\'enic and initially super/trans-Alfv\'enic models. 

The great difference in average f$_{2d}$ that exist between projections for B20 is easily explained considering how the cold gas moves in the three-dimensional box. When $B_0$ is intense the gas's motions are preferentially in the magnetic field's direction producing a more homogeneous distribution in planes parallel to $\vec{B}$, this can be seen in the figures of Appendix~\ref{Angles}. For B20, the density field on YZ is quite homogeneous in space while in the XZ plane the number of lines of sight with cold gas is smaller, this difference has a great influence in the calculations of the mean values which can explain the differences between f$_{yz}$ and f$_{xz}$. Our fractions f$_{yz0}$ and f$_{xz0}$ take a different approach in calculating the average cold mass fraction, as they only consider the beams along which there is cold material, and ignore those with no CNM. The average values of f$_{yz0}$ and f$_{xz0}$, for the same set, are shown in Table~\ref{tab:all_fractions} (fifth and sixth columns). These quantities follow the same trend as the three-dimensional mass fraction, i.e. the average values decrease with an increasing intensity of the $B_0$. The geometry of the density field is less influential in f$_{yz0}$ and f$_{xz0}$ compared to f$_{yz}$ and f$_{xz}$, as can be seen in the smaller differences in the average values between projections.

\subsubsection{Effects of higher mean density}
\label{sec:fractions_dens_set}
\begin{figure}
	\includegraphics[width=\columnwidth]{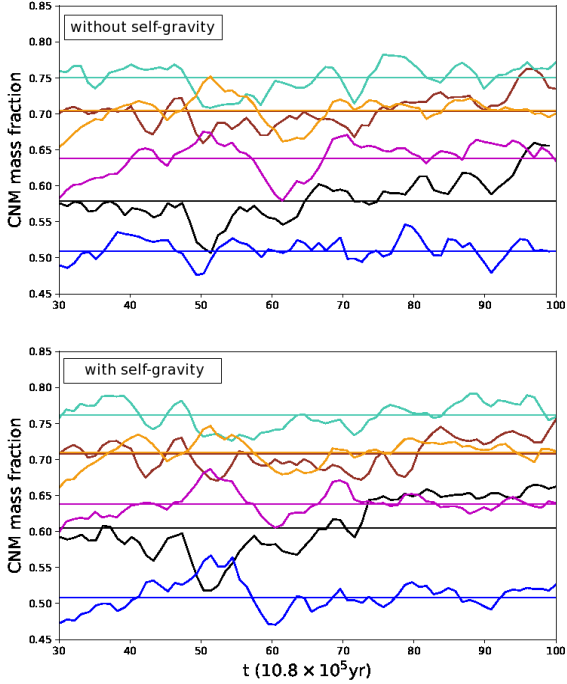}
    \caption{Evolution of the cold gas mass fraction resulting from the three-dimensional data of the high density set and their main set counterpart, without self-gravity (top) and with self-gravity (bottom). The straight lines represent the average values, with the blue ($B_0 \sim 4$\,$\mu$G) and black ($B_0 = 0$\,$\mu$G) lines are for n$_0 = 2$\,cm$^{-3}$, the magenta ($B_0 \sim 4$\,$\mu$G) and brown ($B_0 = 0$\,$\mu$G) lines for n$_0 = 3$\,cm$^{-3}$ and the orange ($B_0 \sim 4$\,$\mu$G) and cyan ($B_0 = 0$\,$\mu$G) ones for n$_0 = 4$\,cm$^{-3}$.}
    \label{fig:3D_fraction_dense}
\end{figure}

The analysis of the resulting CNM mass fractions for the high density set reveals higher f$_{3d}$s as we increase $n_0$, however, the difference between self-gravitating and non self-gravitating pairs is still minimum, as for their main set counterpart. These results can be seen in Fig.~\ref{fig:3D_fraction_dense}. An extremely rough estimate of the amount of cold gas mass segregated, in a closed box, due to TI can be made considering a perfect isobaric ideal segregation (no unstable gas)  as was estimated with eq. (4) in \cite{US}. In Table\,\ref{table3} we show the ideal fractions obtained for our three initial densities. Even if this estimate is crude it does indicate that the cold gas mass fraction should increase as we increase the initial density of our box. The average fractions for the denser simulations are in Table~\ref{tab:all_fractions_dense}. As expected the two-dimensional fractions increase as $n_0$ increases. The effects of self-gravity on these 2D fractions have two scenarios when varying density. First, when B$_0 = 0$ self-gravity lowers the 2D fractions, f$_{2d}$ and f$_{2d0}$, and this gets amplified as $n_0$ increases. Secondly, when B$_0 \sim 4$\,$\mu$G the differences between self-gravitating and non self-gravitating mean 2D fractions are minimal.

\begin{table}
	\centering
	\caption{Average CNM mass fractions for all the simulations from the main set, in three dimensions and both two-dimensional projections.}
	\label{tab:all_fractions}
	\begin{tabular}{lccccr}
        \hline
		Sim. & f$_{3d}$ & f$_{yz}$ & f$_{xz}$ & f$_{yz0}$ & f$_{xz0}$ \\
		\hline
        B00S & 0.5852 & 0.3709 & 0.3806 & 0.4572 & 0.4599\\
		B01S & 0.5776 & 0.3415 & 0.3467 & 0.4707 & 0.4661\\
		B05S & 0.5581 & 0.3566 & 0.3375 & 0.4601 & 0.4589\\
		B10S & 0.5148 & 0.4009 & 0.3532 & 0.4464 & 0.4343\\
        B20S & 0.4738 & 0.4260 & 0.3258 & 0.4399 & 0.4113\\
        B00G & 0.6114 & 0.3657 & 0.3818 & 0.4635 & 0.4701\\
		B01G & 0.5901 & 0.3313 & 0.3402 & 0.4669 & 0.4677\\
		B05G & 0.5653 & 0.3617 & 0.3527 & 0.4643 & 0.4649\\
		B10G & 0.5151 & 0.4026 & 0.3611 & 0.4449 & 0.4393\\
        B20G & 0.4550 & 0.4311 & 0.3337 & 0.4468 & 0.4183 \\
		\hline
	\end{tabular}
\end{table}

\begin{table}
	\centering
	\caption{Cold gas mass fraction segregated in an ideal isobaric scenario.}
	\label{table3}
	\begin{tabular}{lccccr}
        \hline
		$n_0$& Ideal f$_{3d}$\\
		\hline
        cm${^-3}$ & \\
        \hline
        2 & 0.89\\
        3 & 0.93\\
        4 & 0.97\\
		\hline
	\end{tabular}
\end{table}

\begin{table}
	\centering
	\caption{Average CNM mass fractions for all the simulations from the high density set, in three dimensions and both two-dimensional projections.}
	\label{tab:all_fractions_dense}
	\begin{tabular}{lccccr}
        \hline
		Sim. & f$_{3d}$ & f$_{yz}$ & f$_{xz}$ & f$_{yz0}$ & f$_{xz0}$ \\
		\hline
        B00Sn3 & 0.7039 & 0.4933 & 0.4882 & 0.5513 & 0.5503\\
        B00Sn4 & 0.7501 & 0.5717 & 0.5607 & 0.6053 & 0.6159\\
		B10Sn3 & 0.6385 & 0.4989 & 0.4494 & 0.5335 & 0.5184\\
		B10Sn4 & 0.7042 & 0.5665 & 0.5274 & 0.5884 & 0.5703\\
        B00Gn3 & 0.7071 & 0.4798 & 0.4721 & 0.5331 & 0.5321\\
        B00Gn4 & 0.7615 & 0.5253 & 0.5560 & 0.5735 & 0.5846\\
		B10Gn3 & 0.6384 & 0.5089 & 0.4612 & 0.5343 & 0.5174\\
		B10Gn4 & 0.7038 & 0.5675 & 0.5244 & 0.5887 & 0.5699\\
		\hline
	\end{tabular}
\end{table}

\subsection{Alignment of density structures with $\vec{B}$}

The fact that we have different cold gas mass fractions between different projections suggests a geometrical relationship between the magnetic field and the cold gas. This effect seems to be amplified as we increase the value of $B_0$, as can be seen from the 2D fractions presented in section~\ref{sec:fractions_main_set}. Qualitative behaviour of cold gas structures in different projections is described in Appendix~\ref{Angles}. As in the previous section, here we will analyse first the main set, the simulations with mean conditions similar to those of the solar neighbourhood, and then we will examine the effects of higher initial density.

\subsubsection{Main set: effects of varying $B_0$}

A useful procedure to study the relative orientation between the magnetic field and the density structures resulting from simulations is to measure the angle between the density gradient and the magnetic field at every point thus obtaining histograms of relative orientation \citep[HRO, see][]{Soler2013}. As the density gradient points towards the direction of maximum change in density, an isolated density structure will always have gradient vectors normal to its surface. Note however that, as argued in \cite{Clark2014}, the density gradient could be difficult to use in order to characterize observed HI structures because of the low contrast present in this kind of gas. Nonetheless, for numerical simulations aimed to asses the statistical effects of magnetic field, this procedure constitutes a convenient tool.

\begin{figure}
	\includegraphics[width=\columnwidth]{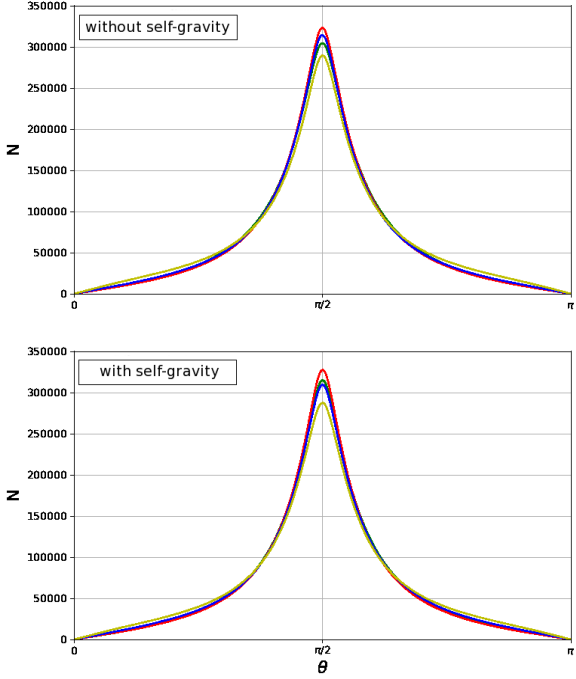}
    \caption{Time averaged histograms of the angle between magnetic field and density gradient for all the gas for simulations without (top) and with (bottom) self-gravity from the main set. The colour code is the same as in Fig.~\ref{fig:3D_fraction}.}
    \label{fig:3D_angles_all}
\end{figure}

The curves in Fig.~\ref{fig:3D_angles_all} are time averaged histograms of the angle $\theta$ between the magnetic field and the density gradient for all the gas in the simulations from our main set. Most of the gas, independently of the initial magnetic field, has a preferred angle at $\theta \sim \frac{\pi}{2}$, implying that the density structures are mainly parallel to the magnetic field. This is independent of the presence of self-gravity.

\begin{figure}
	\includegraphics[width=\columnwidth]{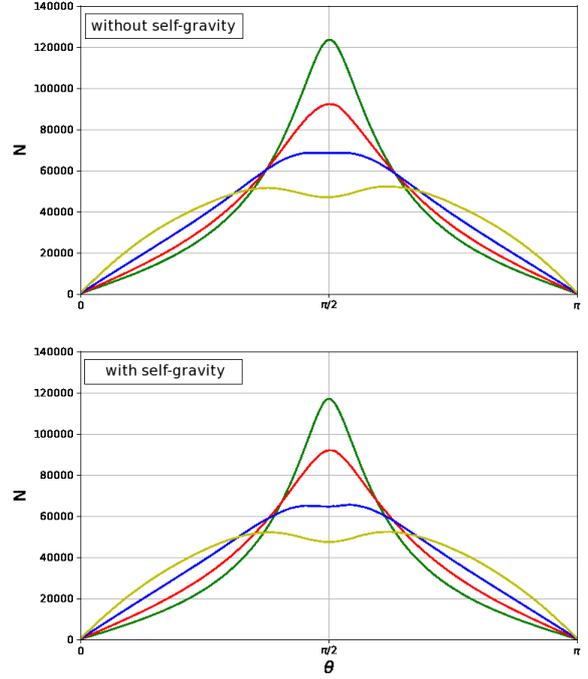}
    \caption{Time averaged histograms of the angle between magnetic field and density gradient for the cold gas for simulations without (top) and with (bottom) self-gravity from the main set. The colour code is the same as in Fig.~\ref{fig:3D_fraction}.}
    \label{fig:3D_angles_cnm}
\end{figure}

A natural question at this point is whether or not the histograms in Fig.\ref{fig:3D_angles_all} are representative of the cold gas structures, which occupy only a small volume fraction in our simulations. In Fig.~\ref{fig:3D_angles_cnm} we show the time averaged histograms of the angle $\theta$ between the magnetic field and the density gradient for the cold gas in the main set. The mayor difference between these histograms and those for the cold gas is that the angle $\theta \sim \frac{\pi}{2}$ is not equally favoured for all the values of $B_0$ that we use. Two different groups can be distinguished: The narrower histograms, with and without self-gravity, result from simulations with weak initial magnetic field, while the wider ones come from the models with high $B_0$. With an increasing magnitude of the initial magnetic field, a dent gradually forms around $\theta \sim \frac{\pi}{2}$, indicating less alignment between the field and the cold dense structures for simulations with high $B_0$.

Even when the three-dimensional data reflect more directly the physical consequences of the variations in the initial conditions for the development of the TI, two-dimensional projections are closer to what an observation may show. In addition, the selection of different elements of the gas helps to emulate observations of cold neutral gas. Finally, working with the most extreme angles of projection (with respect to $B_0$) allows us to understand how meaningful is the projection effect with respect to the values observed in three dimensions.

\begin{figure*}
	\centerline{\includegraphics[width=\textwidth]{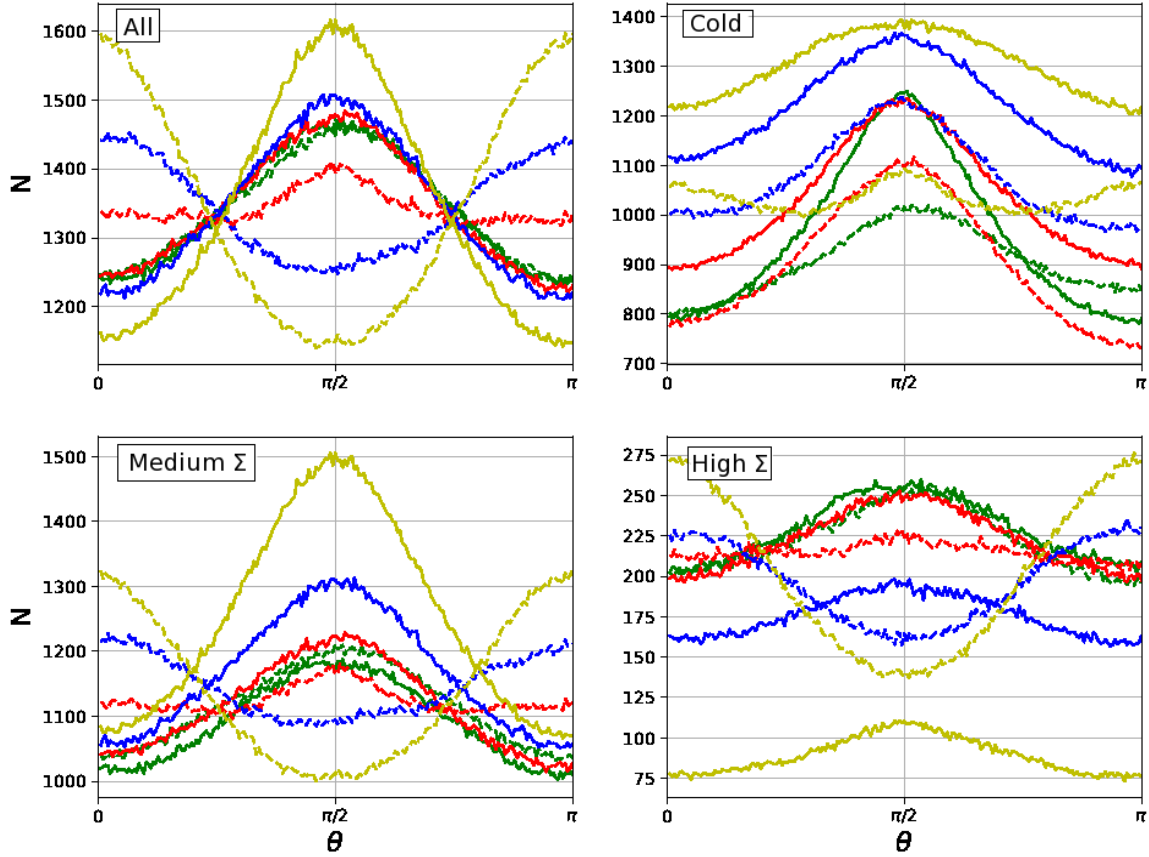}}
    \caption{Histograms of relative angles between the density gradient and the magnetic field for the two-dimensional projections from the main set simulations. The colour scheme is the same as in Fig.~\ref{fig:energiesS}. The continuous line represents the YZ plane and the dashed line the XZ plane. Different panels contain histograms for all the gas in the box (top right), cold gas (top left), directions with moderate $\Sigma$ (bottom right), and directions with high $\Sigma$ (bottom left).}
    \label{fig:2D_anglesS}
\end{figure*}

In order to quantify the relative orientation between the projected cold dense structures and the projected magnetic field (volume-weighted averages) we compute the time averaged HRO for the two projections described above (YZ, XZ). The angle distribution in the two dimensional projections for non self-gravitating runs from the main set can be seen in Fig.~\ref{fig:2D_anglesS}, where we show four panels called All, Cold, Medium $\Sigma$ and High $\Sigma$. The difference between them is the criterion used to select the spaxels considered along each direction. The first panel, All, accounts the integrated density and magnetic field along the direction normal to either the YZ or the XZ plane. Similarly, Cold integrates the spaxels with temperature smaller than or equal to the cold high limit temperature, along the same directions. Two sub groups were taken from All; Medium $\Sigma$, for which we select the pixels where the column density lies between the values of $10^{20}$\,cm$^{-2}$ and $10^{21}$\,cm$^{-2}$, and High $\Sigma$, which includes the pixels where the column density lies between the values of $10^{21}$\,cm$^{-2}$ and $10^{22}$\,cm$^{-2}$.

For All the angle distributions corresponding to YZ (continuous line), look similar to a Gaussian for all $B_0$ cases, while the angle distributions for XZ (dashed line) break this pattern as $B_0$ becomes larger. In fact, for both B10 and B20 we obtain roughly sinusoidal distributions. The most noteworthy difference between projections is for B20 because while in YZ there is a clear preference for an angle $\theta \sim \pi/2$, the XZ plane shows the complete opposite behaviour depleting the $\theta \sim \pi/2$ region. On the other hand, for low $B_0$ values, the behaviour of B01 is very similar between planes, while B05 has the same type of preference for $\theta \sim \pi/2$ but in the XZ plane the maximum is lower and less prominent. 

Meanwhile, for Cold on the YZ plane the quite symmetrical distribution resulting from all of the runs gets broadened as $B_0$ increases. For B01 and B05 the distribution seems to remain approximately Gaussian, while for the higher values of $B_0$ it gets flatter. In the XZ plane, the low $B_0$ simulations have qualitatively the same behaviour as in the All case and the cold-YZ case, namely, the histograms show a preference of the magnetic field to be aligned with the density structures. On the other hand, in the simulations with the two larger values of $B_0$, the cold gas has also a preference for this alignment, in clear contrast to what happens when all the gas is considered. The histogram of B20 on XZ plane shows the population of both $\theta \sim 0$ and $\theta \sim \pi$ angles. This can be explained by taking into account the dominnce of $\vec{B_x}$ in this proyection and the presence of dense structures without strong alineation in 3d, which when proyected in 2d lead to density gradients both parallel and perpendicular to $\vec{B_0}$. Note that there is a slight asymmetry for the cold gas of B01 on XZ. This may be due to the formation of a cold dense structure that has a great influence on the total statistics.

For Medium $\Sigma$ both planes show a behaviour quite similar to the one of All. This is meaningful because we are effectively selecting a representative subset of data that has this characteristic behaviour. In other words, Medium $\Sigma$ material could potentially trace the behaviour of All.

Lastly, for High $\Sigma$ on the YZ plane, the distribution resulting from all models is extremely uniform and displays a small preference for the $\theta \sim \pi/2$ angle which seems to be independent of $B_0$. The vertical differences between the distributions resulting from different models show that the number of dense points decreases as $B_0$ increases. For the XZ plane, only B01 shows a preference for $\theta \sim \pi/2$. while for B05 in this plane, there is not an apparent preference for any angle. For B10 and B20 on XZ the same depletion of $\theta \sim \pi/2$ which characterizes the strong field simulations on the XZ plane can be observed.

Here we only display angle histograms for the simulations without self gravity as the gravitating counterparts show a similar behaviour. The main difference is the amplification of the asymmetry in the distribution of cold gas for B01 on the XZ plane. That is the reason why we associate this asymmetry to the presence of individual denser structures whose effect is amplified when the gravitational force is present.

\subsubsection{Effects of higher mean density}

\begin{figure}
	\includegraphics[width=\columnwidth]{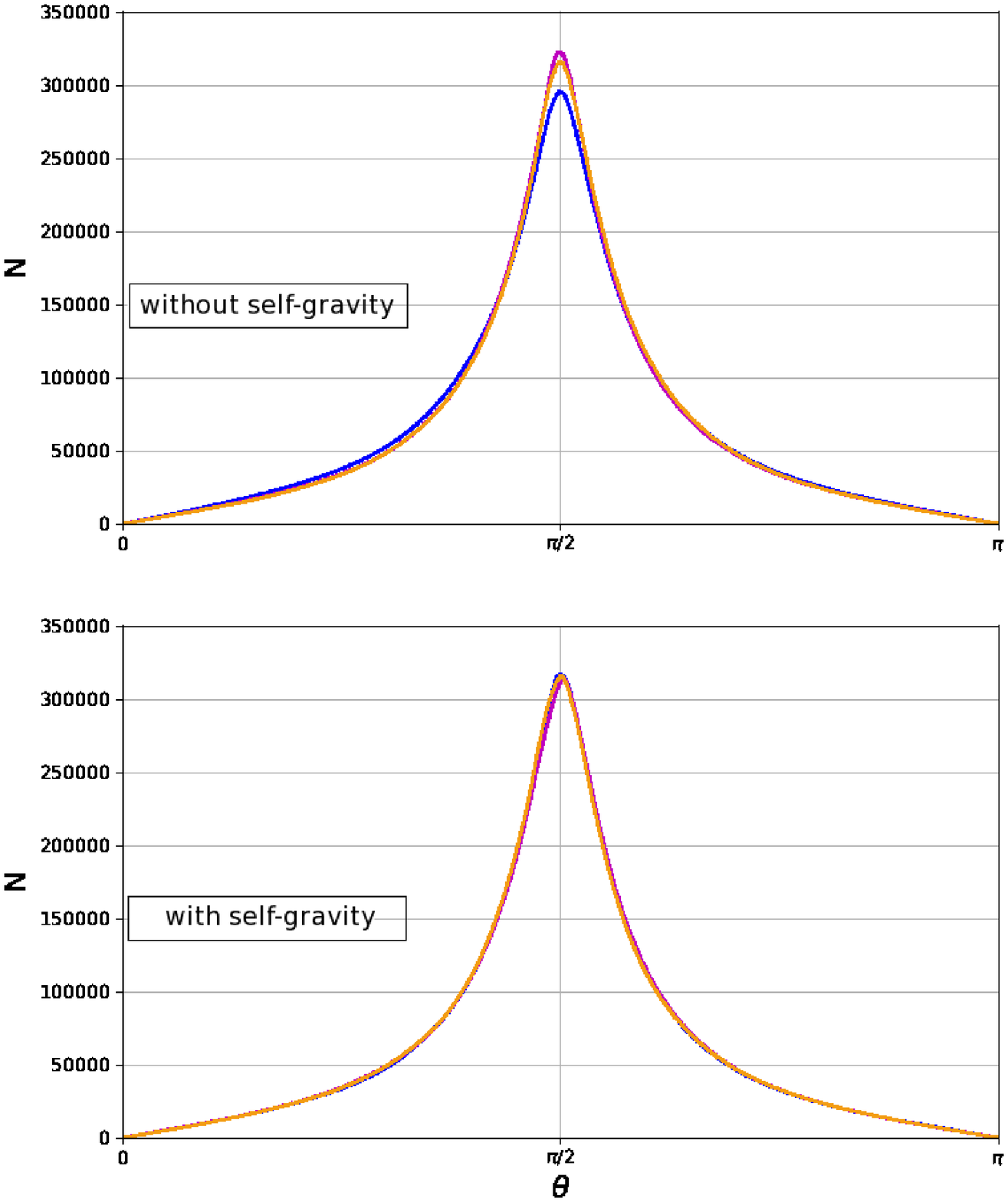}
    \caption{Time averaged histograms of the angle between magnetic field and density gradient for all the gas for simulations without (top) and with (bottom) self-gravity from the high density set and their main set counterpart. The colour code is the same as in Fig.~\ref{fig:3D_fraction_dense}.}
    \label{fig:3D_angles_all_dense}
\end{figure}

\begin{figure}
	\includegraphics[width=\columnwidth]{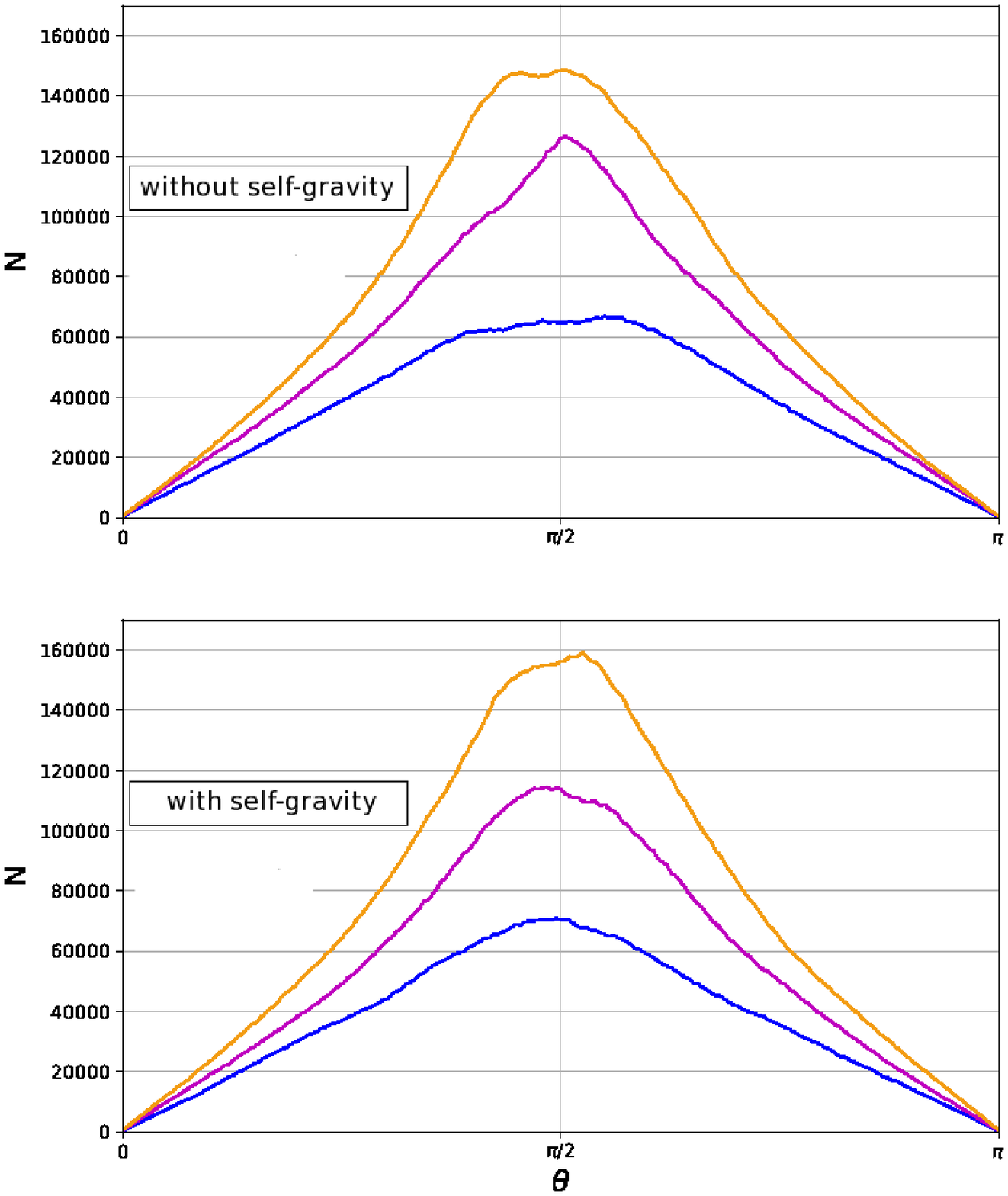}
    \caption{Time averaged histograms of the angle between magnetic field and density gradient for the cold gas for simulations without (top) and with (bottom) self-gravity from the high density set and their main set counterpart. The colour code is the same as in Fig.~\ref{fig:3D_fraction_dense}.}
    \label{fig:3D_angles_cnm_dense}
\end{figure}

\begin{figure*}
	\centerline{\includegraphics[width=\textwidth]{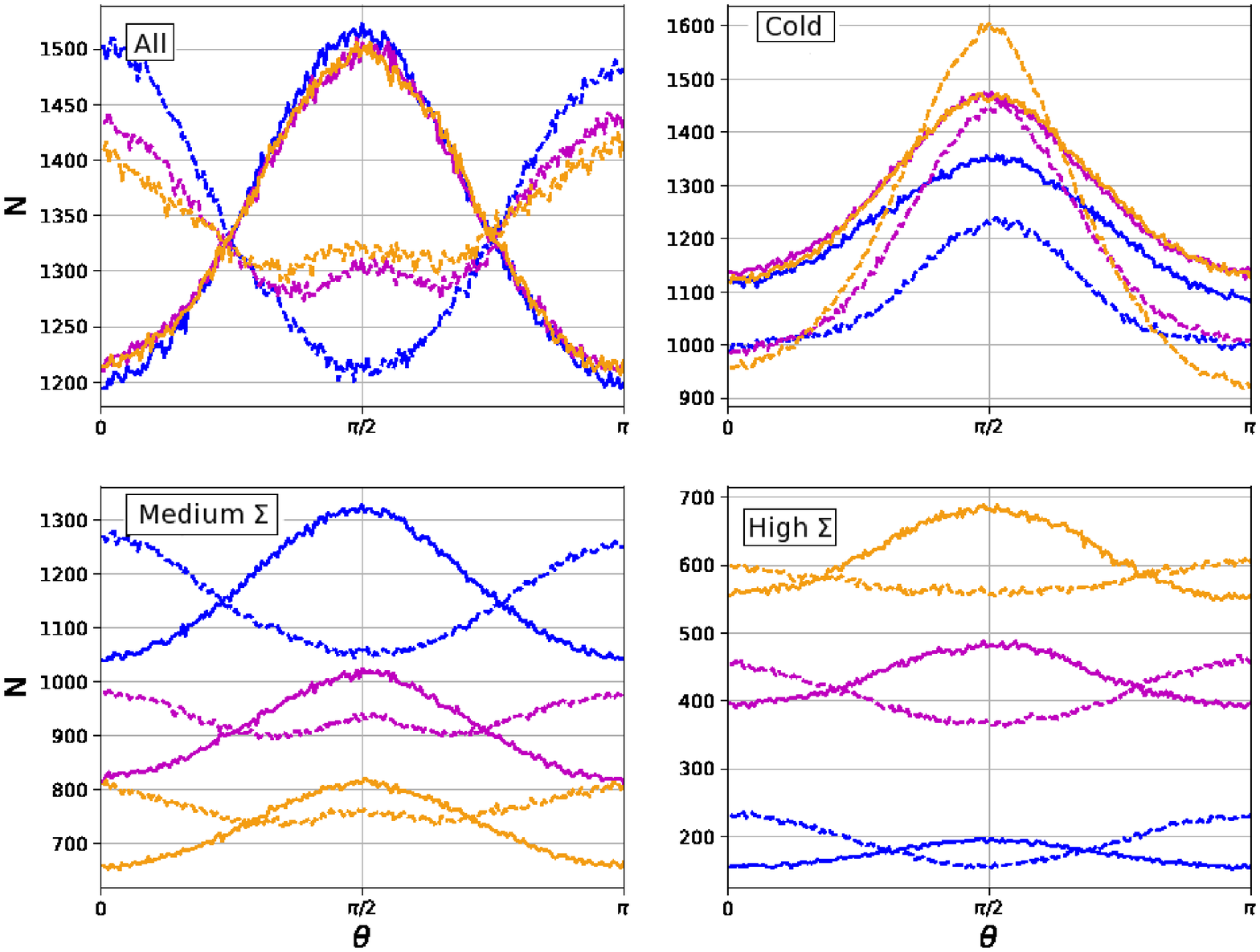}}
    \caption{Histograms of relative angles between the density gradient and the magnetic field for the two-dimensional projections from the high density set and their main set counterpart with self-gravity. The colour scheme is the same as the one used in Fig.~\ref{fig:3D_fraction_dense}. The line style is the same as the one used in Fig.~\ref{fig:2D_anglesS}. Different panels contain histograms for all the gas in the box (top right), cold gas (top left), directions with moderate $\Sigma$ (bottom right), and directions with high $\Sigma$ (bottom left).}
    \label{fig:2D_anglesS_dense}
\end{figure*}

We now proceed to analyse the effects of self-gravity in a medium with higher mean density. In Fig.~\ref{fig:3D_angles_all_dense} we display the histograms for all the gas, which are very similar between themselves without self-gravity (top) and with self-gravity (bottom) and also similar to the ones presented in Fig.~\ref{fig:3D_angles_all}. The histograms shown in Fig.~\ref{fig:3D_angles_cnm_dense} correspond to the CNM of the high density set. As mentioned in section~\ref{sec:fractions_dens_set}, there is more CNM formed in these denser simulations, hence it is seen in the histograms that the number of aligned cold points increases with initial density. The shape of the histograms for B10n3 and B10n4 is, however, different from that resulting from B10n2, the histograms for the denser simulations look more like the ones with lower $B_0$ from the main set. Nevertheless, there are no differences between self-gravitating and non self-gravitating cases. Finally, in Fig.~\ref{fig:2D_anglesS_dense} we can see the effects of having higher mean density in self-gravitating 2D projections. In the first panel we can see a clear preference for the $\theta \sim \pi/2$ angle in the YZ plane. The XZ projection reveals differences in the HROs for all the gas resulting from denser simulations, the angle $\theta \sim \pi/2$ gets more populated as $n_0$ increases. All the histograms in the Cold panel maintain the trends we see in B10n2 but escalated, when a higher $n_0$ is considered, to an environment with more CNM. The third panel, Medium $\Sigma$, seems again to be a good tracer for All, as its trends are quite similar. In the last panel for High $\Sigma$ the trend of the B10 simulation from the main set is conserved, as mirrored sinusoidal distributions, but the amplitudes increase with initial density. It is worth commenting that the 2D HROs resulting from non self-gravitating simulations, even at higher $n_0$, show a similar behaviour to their self-gravitating counterparts. This drove us to omit showing these histograms. 

\section{Discussion}\label{discussion}

We performed MHD simulations in a box with periodic boundary conditions to study the influence of magnetic fields (ranging from $\sim0.1$\,$\mu$G up to $\sim8$\,$\mu$G) on the segregation of neutral atomic gas due to TI. It has been argued that the contribution of magnetic pressure to the total one in the neutral atomic ISM plays an important role \citep[][]{Cox2005}, in particular, the magnetic pressure could provide a supplementary support in regions of low local thermal pressure. Our parameters choice, for the main set of models (see Table\,~\ref{tab:B0_values}), allows for different regimes concerning the ratio of magnetic to thermal pressure together with thermal and kinetic energies of comparable values. The CNM's Mach numbers (sonic and Alfv\'enic) from simulations that encompass the $B_0 \sim 6$\,$\mu$G given by \cite{Heiles2003}, i.e. $B_0 \sim 4$\,$\mu$G (B10) and $B_0 \sim 8$\,$\mu$G (B20), also encompass the Mach numbers reported in the afore mentioned work. It is important to notice that in our simulations the energies attain a stationary regime lasting a period long enough to allow for meaningful time averages. We have also compared the effects of $\vec{B}$ between non-self-gravitating and self-gravitating simulations. In the self gravitating case, magnetic pressure could help counter the effects of gravity in the denser CNM regions. Additionally we analysed simulations with higher mean density, but still in the thermally unstable regime and with values close to those of the solar neighbourhood. This has been done in order to study whether or not an increase of the mass contained in our box could qualitatively affect the behaviour of simulations including self-gravity.  

The first test we used to quantify the influence of $\vec{B}$ in the gas segregation has been to measure the mass fraction of the cold neutral medium. We computed this fraction both, using three-dimensional data (f$_{3d}$), and in two dimensions by averaging over the directions on two planes making different angles with the initial magnetic field (f$_{2d}$ and f$_{2d0}$). The f$_{3d}$ diminishes with an increasing value of $B_0$. When the initial density is closer to the mean value in the solar neighbourhood ($n_0 = 2$\,cm$^{-3}$), the averaged values that we get for f$_{3d}$ range between $\sim 47$\% and $\sim 58$\%. This range is  slightly above the measured value for the solar neighbourhood, $\sim 40\%$, reported by \cite{Heiles2003} and \cite{Pineda2013}, which is consistent with results from more recent observational studies \citep[][]{Roy2013, Murray2015, Roy2017}. We find that the average value of f$_{2d0}$ tends to decrease as $B_0$ becomes larger, mimicking the behaviour of f$_{3d}$. The f$_{2d0}$ mass fraction's average values do not show a strong sensibility to the plane direction, while average values of f$_{2d}$ are smaller when computed by averaging over directions normal to $\vec{B_0}$. These differences highlight the effects of projection that the geometry of the cold structures has. The high density set ($n_0 = 3$\,cm$^{-3}$ and $n_0 = 4$\,cm$^{-3}$) has f$_{3d}$s ranging between $\sim 64$\% and $\sim 76$\%, values up to $\sim 36\%$ higher than what is observed, this is a direct result of the higher amounts of cold gas formed. Nevertheless the tendency to decrease in average when observed in a projection normal to $\vec{B_0}$ is independent of $n_0$. As expected, the increase in mean density increases the resulting cold gas mass fraction, nevertheless, self-gravity does not have major effects in f$_{3d}$ even with higher $n_0$.

We can compare our f$_{3d}$s with other simulations. One of the numerical works that has also measured cold gas mass fraction is \cite{Hill2012}, who find three-dimensional fractions rounding $\sim 60\%$ for SN driven MHD simulations with vertical disk structure, agreeing both with the results presented in \cite{US}, obtained from hydrodynamical simulations in a periodic box, and with the ones reported in the present work. The recent values reported in \cite{Pardi2017}, with periodic simulations with SN driven turbulence,  are quite different from the ones we obtain, rounding $\sim 90\%$. This could be due to the differences in resolution ($\sim 4$\,pc) and energy injection scheme  used in their models. Nevertheless, our results are in agreement with both aforementioned MHD works concerning the fact that the increase in initial magnetic field intensity diminishes the mass fraction of cold gas. This is opposite to what \cite{Heitsch2009} find using simulations of non self-gravitating colliding flows in the case where the flows are moving along the magnetic field, this could indicate that a preferential angle between $\vec{B}_0$ and the bulk velocity affects the statistical thermal properties of the gas.

Another outcome of this analysis is that, at average solar neighbourhood conditions, the CNM segregation is more influenced by a varying the intensity of $\vec{B}$  than by varying $v_{rms}$. This fact could be relevant to explain the behaviour of the cold gas mass fraction at outer Galactic radii which has been reported as approximately constant beyond the radii $R \sim 11$\,kpc \citep[][]{Dickey2009, Pineda2013}. An attempt on explaining this, in terms of $v_{rms}$ variations, is presented in \cite{US}, however the current finding suggest that the inclusion of magnetic fields in simulations done with the conditions of different galactocentric radii should also considerably change the amount of segregated CNM.

In our main set models ($n_0 = 2$\,cm$^{-3}$), the influence of self-gravity on the CNM fraction varies depending on the initial magnetic field. The purely hydrodynamical and the weakest initial field (B01: $B_{0} \sim 0.4$\,$\mu$G) simulations get their f$_{3d}$ significantly raised with the inclusion of self-gravity. A distinct behaviour is observed in the simulations B05 ($B_0 \sim 2$\,$\mu$G) and B10 ($B_0 \sim 4$\,$\mu$G), which do not show great differences between self-gravitating and non self-gravitating runs. The same is true for models with $B_0 \sim 4$\,$\mu$G and higher initial density ($n_0 = 3$\,cm$^{-3}$ and $n_0 = 4$\,cm$^{-3}$). Finally, the simulation with the strongest initial field (B20: $B_{0} \sim 8$\,$\mu$G) shows a smaller f$_{3d}$ when self-gravity is present. We attribute this decrement to the fact that a strong magnetic field partially inhibits the TI and prevents the formation of large enough high-density regions, where gravity is more influential. The figures in Appendix~\ref{Angles} suggest that a statistical analysis on the physical properties of individual cold dense structures is needed to get more insight on this interpretation, however such study is beyond the scope of the present work. For the mass fractions obtained from two-dimensional projections, the effects of self gravity are not as noticeable, this could be due to our way of obtaining these quantities, which involves averaging twice: measuring the CNM mass fraction along each line of sight and then averaging over those fractions.

The interplay between TI and $\vec{B}$ has been recently studied in the context of the formation of cold neutral and molecular gas in a shock front set-up where there is a range of angles between magnetic field and the shock front. \citep[][]{Banerjee2015, Fogerty2016, Inoue2016}. In \cite{Banerjee2015} the authors report a delay in the condensation of cold gas as  the angle between $\vec{B}$ and the shock front is increased. They also find that this effect is amplified as they increase the intensity of $\vec{B}$. More recently, \cite{Fogerty2016} and \cite{Inoue2016}, with a similar set-up, arrived at the same conclusions. The transition from WNM to CNM via TI induced by strong compressions has been recognized as a precursor of the formation of molecular clouds \citep[see e.g.][for a comprehensive review of this topic see \cite{Enrique2015}.]{Hennebelle1999, Koyama2000, Enrique2006, Heitsch2008}. With this in mind, the amount of cold neutral gas segregated should influence the amount of star forming clouds and the gas reservoir for star formation, hence having a magnetic field lowering the CNM fraction could lower the amount of stars formed. All of this is consistent with the diminishing of $f_{3d}$ as $B_0$ increases, which we attribute to the partial inhibition of the TI predicted by \cite{Field65}. This inhibition could also be related with the lowering of the star formation rate (SFR) and the amount of star forming regions as reported by \cite{Iffrig2014} for disk simulations with SN feedback. However this has also been observed in isothermal simulations, \citep[see for example ][]{Price2008, Dib2010, Padoan2011}, indicating that the phenomenon is at least in part due to the influence of magnetic pressure in the collapsing regions. A possible way to quantify the relative contribution of these mechanisms to the lowering of the SFR could be to study the evolution of dense gas obtained from simulations like ours, but with decaying turbulence, however, that kind of study is beyond the scope of the present work.

For the second half of our study we looked for the relative angles that exist between the magnetic field and the dense structures, in three and two dimensions through the HRO. The three-dimensional analysis of the relative angles between magnetic field and density gradient being independent of projection effects, shows the actual alignments between these vectors. For our models B01, B05, and even B10 we find a true alignment between the local magnetic field and the cold structures independently of the presence of self-gravity and also independently of the mean density of our computational box (B10n3 and B10n4). In fact, for our initial fields B01 and B05 the alignment is so strong that it can also be observed in the histograms obtained from the two-dimensional projections, independently of how extreme the projection angle is with respect to the initial orientation of $\vec{B}$ (parallel and perpendicular), and also independently of which subgroup of the data we use for the analysis (cold gas, medium or high column density points). The alignment for the three mean densities tested with $B_0 \sim 4$\,$\mu$G is strong in the cold gas independently of projection. For All, Medium $\Sigma$ and High $\Sigma$ this alignment gets lost when $n_0 = 2$\,cm$^{-3}$, but with higher densities it is clear that a preferred alignment is starting to appear and it mildly intensifies with density. The resulting alignment between cold structures and the local magnetic field is in agreement with recent observational results. On one hand, in the work of \cite{Clark2014}, the authors report an alignment between the local magnetic field and the cold structure in observations of an HI cloud with an unknown angle of projection with respect to the local $\vec{B}$, even if the scatter is significant. Nevertheless we should note that the structures described in the aforementioned work are CNM fibres with $\sim 0.01$\,pc width, hence smaller than the structures analysed here. On the other hand, in \cite{PlanckXXXII}, the authors find an alignment between the diffuse observed structures and the magnetic field. It is worth noticing that they interpret their results as a possible consequence of the presence of super-Alfv\'enic turbulence. From the numerical perspective, our results are complementary and in agreement with the results of \cite{Inoue2016}, who analysed shock compressed layer simulations with background $\vec{B}$ (and different angles between it and the shock front), and found a predilection for an alignment between CNM structures and magnetic field with variations just in the scatter of the data.

An effect that could possibly have an important role in the alignment of the dense structures with $\vec{B}$ is the presence of thermal conductivity, not included in our models, which can be particularly relevant in magnetized cases. It is known that there is a critical length, named Field length \citep[see ][]{Begelman1990}, below which TI is inhibited by thermal conductivity \citep[][]{Field65}. This length is estimated to be $\sim 0.1$\,pc for the WNM and $\sim 1 \times 10^{-3}$\,pc for the CNM \citep[][]{Audit2005}. In numerical simulations, limited resolution has an equivalent effect when the smallest resolved scale is larger than Field's length \citep[][]{Gazol2005}. When magnetic field is present, thermal conduction becomes anisotropic due to the preferential direction of electron motion along magnetic field, which results on a reduction of thermal conductivity across $\vec{B}$. The effect of this anisotropy on the non-linear development of TI has been studied by \cite{Choi2012}, who concluded that the statistics of density and temperature are not strongly affected by anisotropic conduction, but the geometry and orientation (with respect to $\vec{B}$) of the structures should be affected. As the dense structures would be elongated along $\vec{B}$, the alignment between them and the field could be even stronger than what we have measured. The alignment between dense structures and the field in B20 could then increase with the inclusion of thermal conductivity. This is because increasing the intensity of $\vec{B}$ increases the conductivity's degree of anisotropy and that this could increase the alignment between dense structures and $\vec{B}$.


\section{Conclusion}\label{conclusion}

In this paper we have studied the influence of magnetic fields on the CNM using a set of MHD simulations with continuous Fourier forcing, different initial magnetic fields and in some cases self-gravity. In particular, we analysed how $\vec{B}$ affects the segregation of cold gas, measured by its mass fraction, and what is the alignment between dense structures and this field. We have also studied the effect of the presence of self-gravity on both quantities. From the results presented in previous sections we arrive at the following conclusions:

\begin{itemize}
 
\item The influence of the magnetic field intensity over the mass fraction of segregated CNM is quite relevant and, for solar neighbourhood conditions, varying its initial value affects the segregation more than varying the $v_{rms}$.

\item When there are no imposed orientations between the initial magnetic field and the bulk velocity, the CNM mass fraction is lowered by the presence of $\vec{B}$ and it reaches lower values as the magnetic field intensity increases. 

\item The mass fraction measured as an average along directions with fixed orientation with respect to $\vec{B_0}$ differs between projection. This suggests that the geometry of the cold structures is heavily influenced by the magnetic field. 

\item Cold gas dense structures tend to be aligned with the local magnetic field. This is prevalent in all the studied cases, but the increase in intensity of $\vec{B_0}$ diminishes the degree of alignment. This tendencies are also observed, when $\vec{B_0}$ is weak, in two-dimensional projections both in directions parallel and perpendicular to the initial magnetic field. 

\item The presence of self-gravity does not affect significantly the behaviour of the CNM's mass fraction nor the alignment of cold structures with $\vec{B}$ when using initial mean densities akin to those of the solar neighbourhood. Given a fixed initial magnetic field but using an increased, up to double, mean density there are not qualitative changes with the addition of self-gravity. This could change in a case of decaying turbulence or in models of collapsing regions.

\end{itemize}


\section*{Acknowledgments}

M. A. Villagran acknowledges support from a CONACYT scholarship.
Both authors appreciate the support of  UNAM-DGAPA, PAPIIT through the grant: IN110316. This work has made extensive use of the NASA's Astrophysics Data System Abstract Service. We also thank the anonymous referee for the suggestions that helped improve our work. 



\bibliographystyle{mnras}
\bibliography{references} 



\appendix

\section{Column density maps}\label{Angles}

For a better understanding of the spatial distribution of the cold gas in our simulations, we present here a qualitative description of the cold structures via projected density and $\vec{B}$ maps. All of the figures were done at a time $t \sim 8.6 \times 10^6$\,yr, and the scalar fields are the column numerical densities of the cold gas, while the arrows represent the direction of the projected magnetic field associated to the cold gas.

\begin{figure*}
	\centerline{\includegraphics[width=\textwidth]{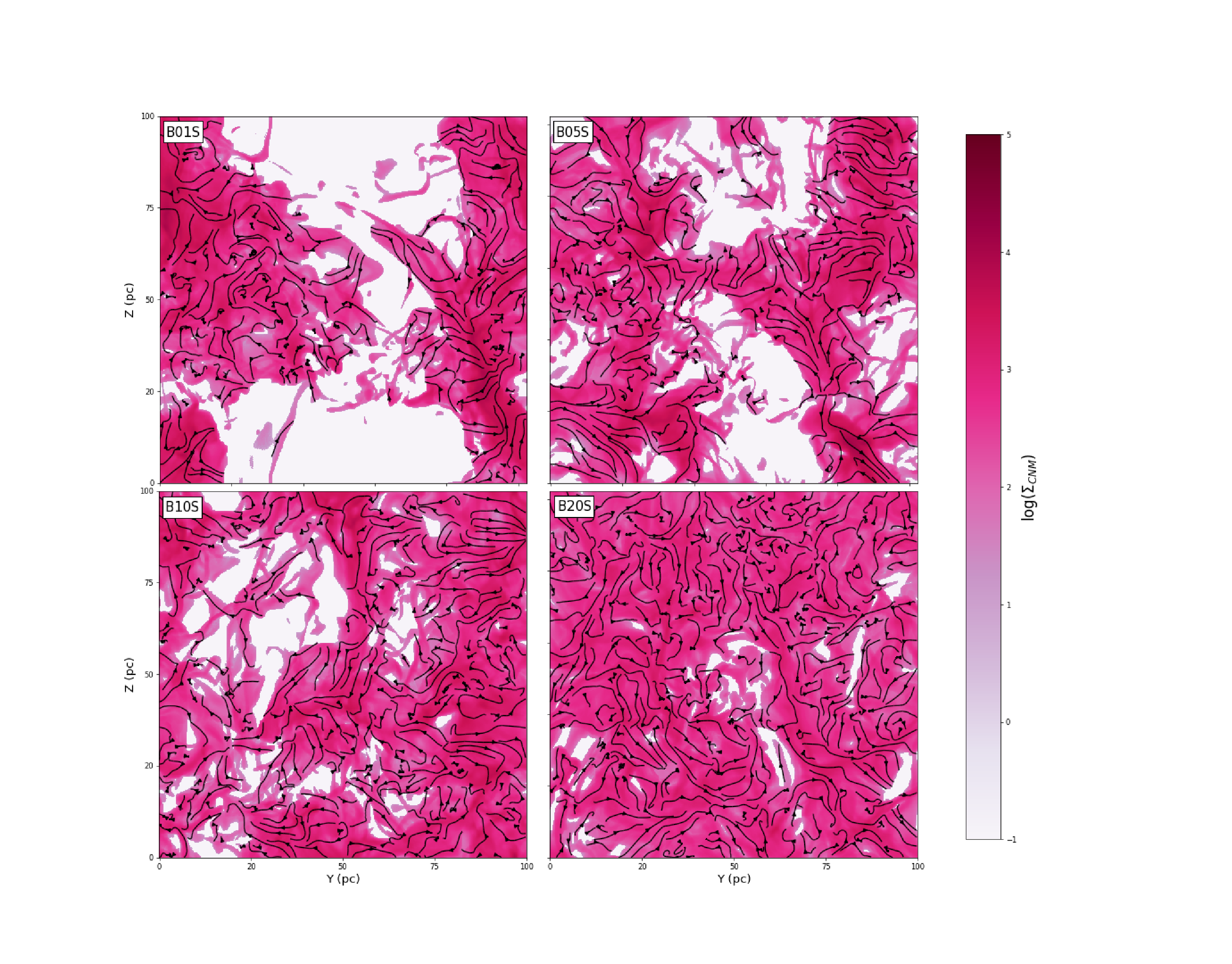}}
    \caption{Projected density and magnetic field on the YZ plane for the simulations without self-gravity.}
    \label{fig:laminas_YZ_T80S}
\end{figure*}

The four panels in Fig.~\ref{fig:laminas_YZ_T80S} are each a projection on the YZ plane of the magnetized simulations without self-gravity. It can be observed that while there are dense structures in all of the panels, the size of these structures decreases with $B_0$. In this projection the magnetic field seems to distribute the cold gas around the plane. With the strongest $B_0$, the resulting density field is more homogeneous while the weakest $B_0$ gives some strongly concentrated regions. There is not an obvious relationship between the magnetic field lines and the dense structures, even for the strongest initial field the lines seem uncorrelated with most of the structures. Some of the dense regions are clearly surrounded by field lines, the most noteworthy ones in B05S, but it is hard to establish a strong correlation.

\begin{figure*}
	\centerline{\includegraphics[width=\textwidth]{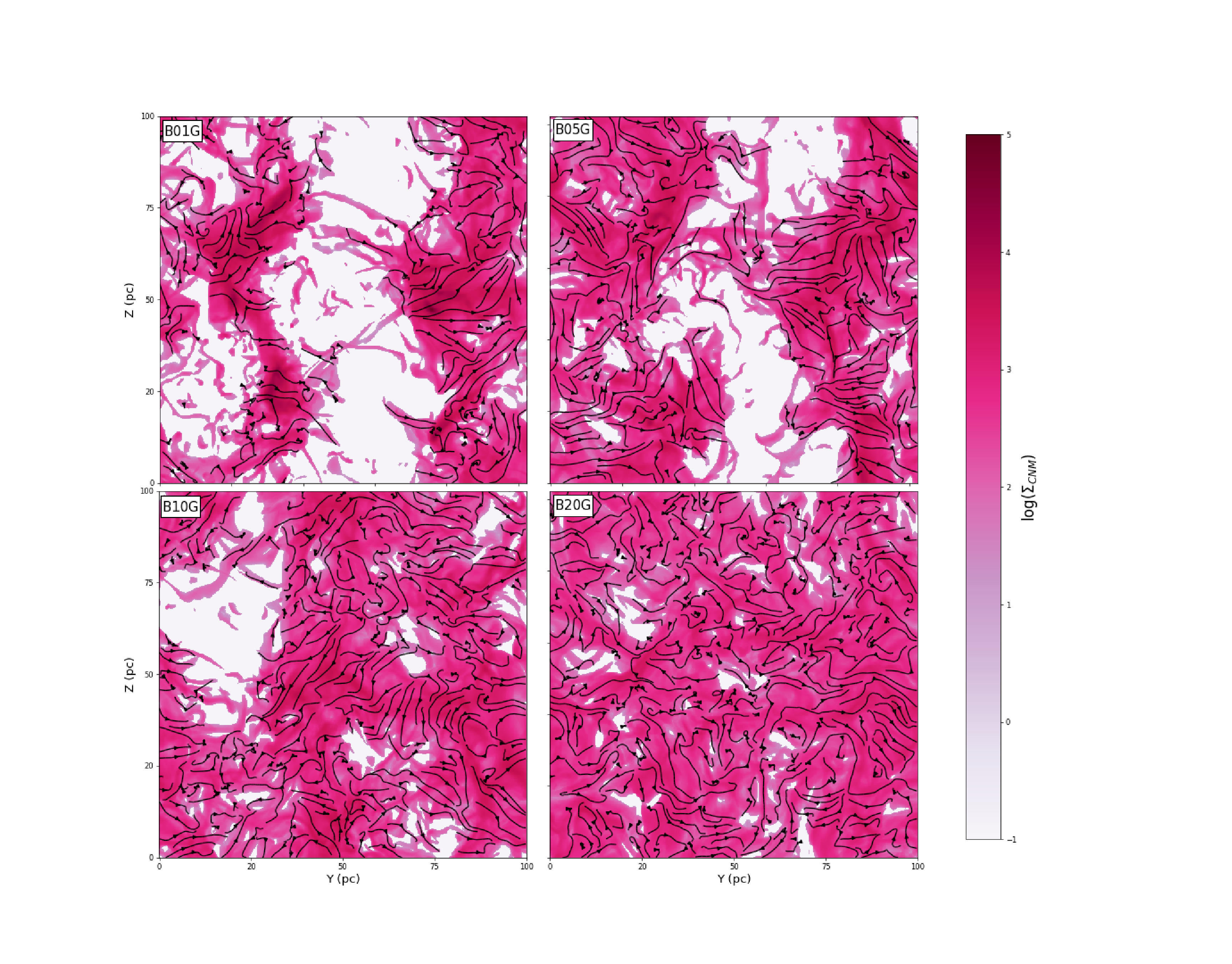}}
    \caption{Projected density and magnetic field on the YZ plane for the simulations with self-gravity.}
    \label{fig:laminas_YZ_T80G}
\end{figure*}

In Fig.~\ref{fig:laminas_YZ_T80G} we show the same projections but for the self-gravitating simulations. A first observation is that the dense structures are more concentrated even for the relatively high initial magnetic field of B10G, but the homogenizing effect of the magnetic field is still present. It is thus clear from this plots that the effect of adding self-gravity at these density ranges is not altering the dynamics in a significant way.

\begin{figure*}
	\centerline{\includegraphics[width=\textwidth]{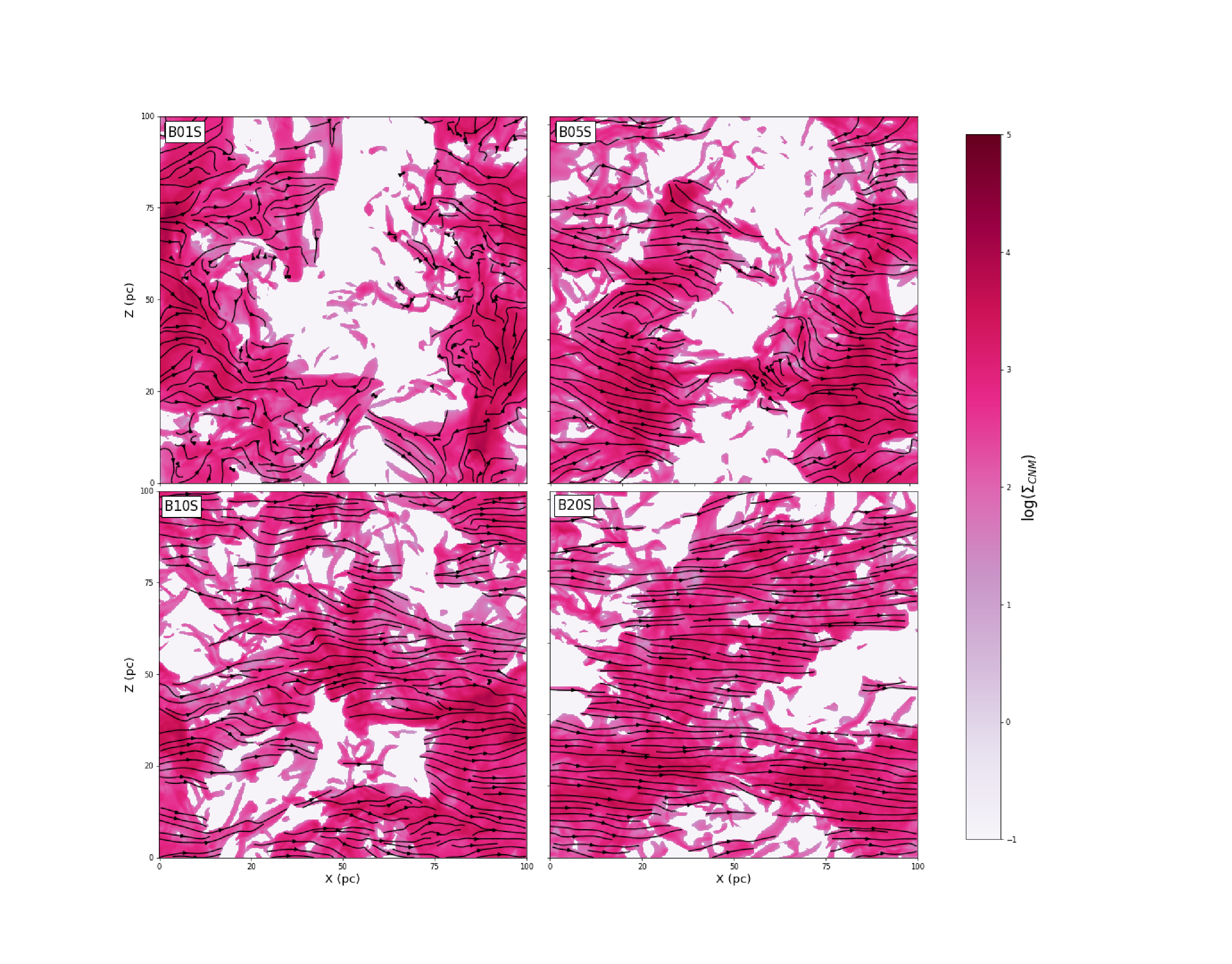}}
    \caption{Projected density and magnetic field on the XZ plane for the simulations without self-gravity.}
    \label{fig:laminas_XZ_T80S}
\end{figure*}

Next we explore what happens when the projection direction is normal to the initial magnetic field. In Fig.~\ref{fig:laminas_XZ_T80S} we display the projection onto the XZ plane of the simulations without self-gravity. Here we can see that, as expected, the uniformity of the magnetic field increases with its initial intensity. The dense structures decrease in size the stronger the initial magnetic field is. In this projection, the simulation B01S, with the weakest field, is the only one presenting magnetic field lines with an important curvature near dense structures. The same tendency is lightly seen in B05S near the densest structures but in general the projected field is more uniform. The remaining simulations, with the strongest magnetic fields, have magnetic vectors that preserve their uniformity almost independently of the presence of dense structures.

\begin{figure*}
	\centerline{\includegraphics[width=\textwidth]{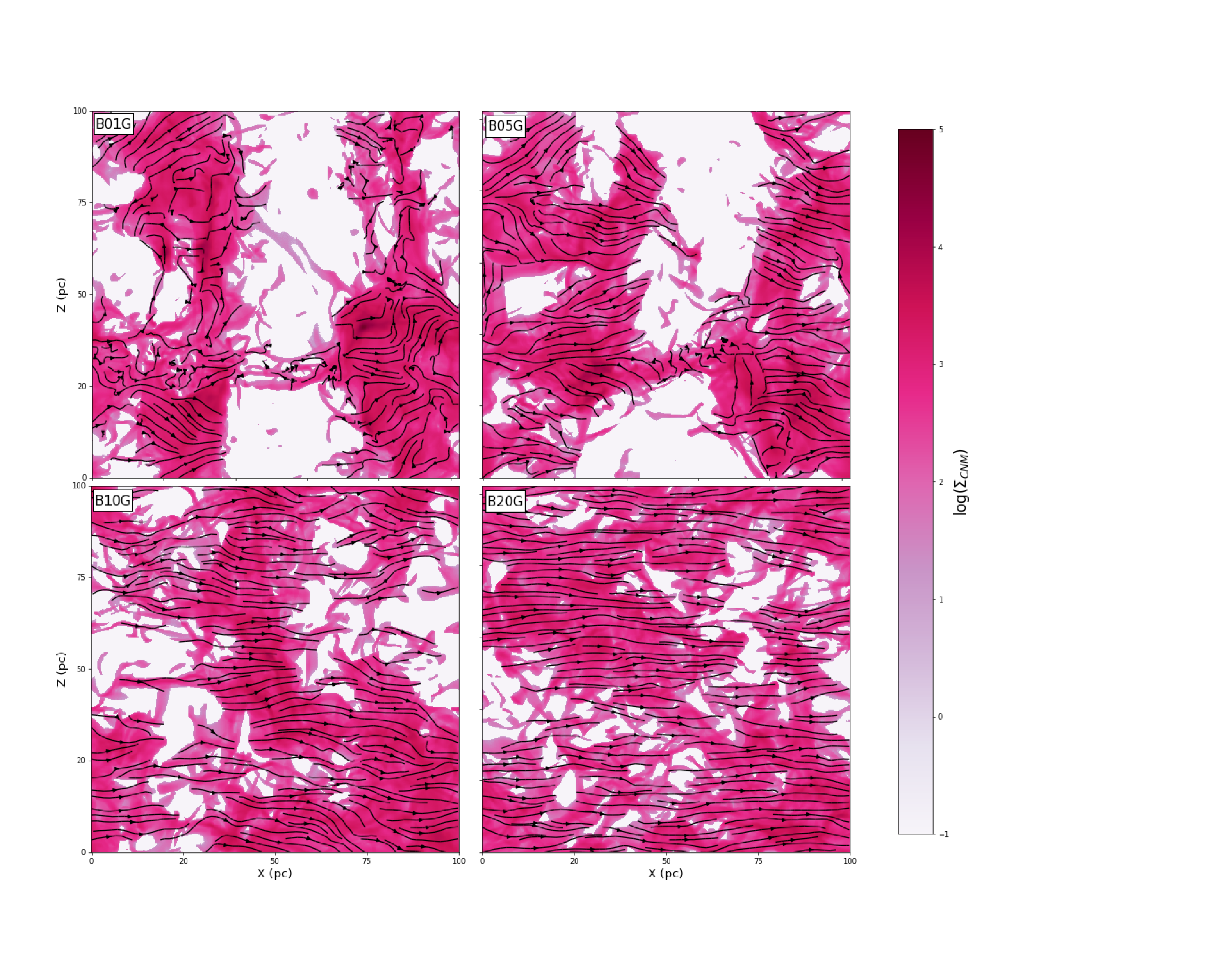}}
    \caption{Projected density and magnetic field on the XZ plane for the simulations with self-gravity.}
    \label{fig:laminas_XZ_T80G}
\end{figure*}

Finally, we show in Fig.~\ref{fig:laminas_XZ_T80G} the same projection for the simulations with self-gravity. In this case, the morphology of the formed dense regions is slightly different than before for all the simulations, having cold structures smaller than in Fig.~\ref{fig:laminas_XZ_T80S} but the trend in the relationship between structure size and intensity of magnetic field is similar. We can also observe more bent field lines here than in the plot for the non-self-gravitating simulations in every initial intensity slot. Curling around dense structures is easily noticeable in the super Alfv\'enic simulations while for the rest the bending is less severe.

\bsp	
\label{lastpage}
\end{document}